\documentclass{ws-ijufks}
\usepackage{epic}
\usepackage{eepicemu}
\usepackage{graphicx}
\usepackage{latexsym}
\usepackage{subfigure}
\usepackage{epsfig}

\begin{document}

\markboth{Rajshekhar Sunderraman and Haibin Wang}{Paraconsistent Intuitionistic Fuzzy Relational Data Model}

\bibliographystyle{splncs}

\catchline{}{}{}{}{}

\title{PARACONSISTENT INTUITIONISTIC FUZZY RELATIONAL DATA MODEL} 

\author{RAJSHEKHAR SUNDERRAMAN and HAIBIN WANG}
\address{Department of Computer Science \\
Georgia State University\\
Atlanta, Georgia 30302, USA \\
raj@cs.gsu.edu, hwang17@student.gsu.edu}

\maketitle

\begin{history}
\received{(received date)}
\revised{(revised date)}
%\accepted{(Day Month Year)}
%\comby{(xxxxxxxxxx)}
\end{history}

\begin{abstract}
In this paper, we present a generalization of the relational data
model based on paraconsistent intuitionistic fuzzy sets.
Our data model is capable of manipulating incomplete as well as
inconsistent information. 
Fuzzy relation or intuitionistic fuzzy relation
can only handle incomplete information.
Associated with each relation are two membership functions one is 
called truth-membership function $T$ which
keeps track of the extent to which we believe the tuple is in the
relation, another is called false-membership function which keeps 
track of the extent to which we believe that it is not in the relation.
A paraconsistent intuitionistic fuzzy relation is inconsistent if there
exists one tuple $a$ such that $T(a) + F(a) > 1$.
In order to handle inconsistent
situation, we propose an operator called split to transform inconsistent
paraconsistent intuitionistic fuzzy relations into pseudo-consistent
paraconsistent intuitionistic fuzzy relations and do the set-theoretic and
relation-theoretic operations on them and finally use another operator called
combine to transform the result back to paraconsistent intuitionistic fuzzy
relation.
For this model, we define algebraic operators that are generalisations
of the usual operators such as union, selection, join on fuzzy
relations.
Our data model can underlie any database and knowledge-base management 
system that deals
with incomplete and inconsistent information.
\end{abstract}


\section{Introduction} \label{Intro}

Relational data model was proposed by Ted Codd's pioneering paper \cite{cdd70}.
Since then, relational database systems have been extensively studied and 
a lot of commercial relational database systems are currently 
available \cite{en2000,sks96}. This data model usually takes care of only
well-defined and unambiguous data. However, imperfect information is 
ubiquitous
-- almost all the information that we have about the real world is not certain,
complete and precise~\cite{prs96}. Imperfect information can be classified as:
incompleteness, imprecision, uncertainty, inconsistency. Incompleteness arises
from the absence of a value, imprecision from the existence of a value which
cannot be measured with suitable precision, uncertainty from the fact that
a person has given a subjective opinion about the truth of a fact which he/she  
does not know for certain, and inconsistency from the fact that there are
two or more conflicting values for a variable.

In order to represent and manipulate various forms of incomplete information in
relational databases, several extensions of the classical relational model have
been proposed \cite{bsk83,bd84,cdd79,lip79,lps:datinc,mai83}. In some of these extensions, a variety of "null values" 
have been introduced to model unknown or not-applicable data values. Attemps 
have also been made to generalize operators of relational algebra to manipulate
such extended data models \cite{bsk83,cdd79,mai83}. The fuzzy set theory and fuzzy logic 
proposed by Zadeh \cite{zdh65} provide a requisite mathematical framework for 
dealing with incomplete and imprecise information. 
 Later on, the concept of
interval-valued fuzzy sets was proposed to capture the fuzziness of grade
of membership itself~\cite{TUR86}. In 1986, Atanassov introduced the
intuitionistic fuzzy set~\cite{ATA86} which is a generalization of fuzzy
set and provably equivalent to interval-valued fuzzy set. The intuitionistic
fuzzy sets consider both truth-membership $T$ and false-membership $F$ with
$T(a), F(a) \in [0, 1]$ and $T(a) + F(a) \leq 1$. Because of the restriction,
the fuzzy set,
interval-valued fuzzy set and
intuitionistic fuzzy set cannot handle inconsistent information.
Some authors \cite{nr84,bld83,bp82,ck78,kz86,prd84,rm88} have
studied relational databases in the light of fuzzy set theory with an objective
to accommodate a wider range of real-world requirements and to provide closer
man-machine interactions. Probability, possibility and Dempster-Shafer theory 
have been proposed to deal with uncertainty. Possibility theory ~\cite{zdh78}
is bulit upon the idea of a fuzzy restriction. That means a variable could only
take its value from some fuzzy set of values and any value within that set is 
a possible value for the variable. Because values have different
degrees of membership in the set, they are possible to different degrees.
Prade and Testemale ~\cite{prt84} initially suggested using 
possibility theory to deal with incomplete and uncertain information in 
database. Their work is extended in ~\cite{prt87} to cover multivalued
attributes. Wong~\cite{wng82} proposes a method that quantifies the uncertainty in a databse using probabilities. His method maybe is the simplest one which attaches a probability to every member of a relation, and to use these values
to provide the probability that a particular value is the correct answer to
a particular query. Carvallo and Pittarelli ~\cite{cvp87}
also use probability theory to model uncertainty in relational databases systems. Their method augmented projection an join operations with probability measures.  

However, unlike incomplete, imprecise and uncertain information, 
inconsistent information has not
enjoyed enough research attention. In fact, inconsistent information exists
in a lot of applications. For example, in data warehousing application, 
inconsistency will appear when trying to integrate the data from many different
sources. Another example is that in the expert system, there exist facts which
are inconsistent with each other. 
Generally, two basic approaches have been
followed in solving the inconsistency problem in knowledge bases: belief revision and paraconsistent logic. The goal of the first approach is to make an 
inconsistent theory consistent, either by revising it or by representing it
by a consistent semantics. On the other hand, the paraconsistent approach
allows reasoning in the presence of inconsistency, and contradictory information can be derived or introduced without trivialization \cite{mcm2002}. 
Bagai and Sunderraman \cite{bgs95a} proposed a paraconsistent
relational data model to deal with incomplete and inconsistent information.
This data model is based on paraconsistent logics which were studied in detail
by de Costa \cite{cst77} and Belnap \cite{bln77}.

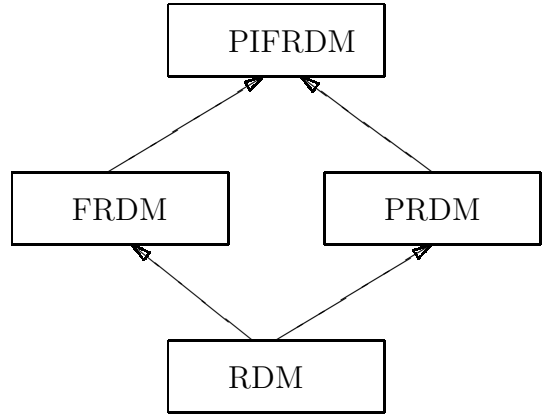
\begin{figure}[htbp]
\begin{center}
%{\includegraphics{fig1.eps}}
\setlength{\unitlength}{0.00083333in}
\begingroup\makeatletter\ifx\SetFigFont\undefined%
\gdef\SetFigFont#1#2#3#4#5{%
  \reset@font\fontsize{#1}{#2pt}%
  \fontfamily{#3}\fontseries{#4}\fontshape{#5}%
  \selectfont}%
\fi\endgroup%
{\renewcommand{\dashlinestretch}{30}
\begin{picture}(3324,2589)(0,-10)
\path(987,2112)(2337,2112)(2337,2562)
	(987,2562)(987,2112)
\path(12,1062)(1362,1062)(1362,1512)
	(12,1512)(12,1062)
\path(1962,1062)(3312,1062)(3312,1512)
	(1962,1512)(1962,1062)
\path(987,12)(2337,12)(2337,462)
	(987,462)(987,12)
\path(1662,462)(2637,1062)
\blacken\path(2550.524,973.559)(2637.000,1062.000)(2519.078,1024.658)(2550.524,973.559)
\path(612,1512)(1587,2112)
\blacken\path(1500.524,2023.559)(1587.000,2112.000)(1469.078,2074.658)(1500.524,2023.559)
\path(2637,1512)(1812,2112)
\blacken\path(1926.693,2065.681)(1812.000,2112.000)(1891.403,2017.157)(1926.693,2065.681)
\path(1512,462)(762,1062)
\blacken\path(874.445,1010.463)(762.000,1062.000)(836.963,963.611)(874.445,1010.463)
\put(1362,2262){\makebox(0,0)[lb]{\smash{{{\SetFigFont{12}{14.4}{\rmdefault}{\mddefault}{\updefault}PIFRDM}}}}}
\put(387,1212){\makebox(0,0)[lb]{\smash{{{\SetFigFont{12}{14.4}{\rmdefault}{\mddefault}{\updefault}FRDM}}}}}
\put(2337,1212){\makebox(0,0)[lb]{\smash{{{\SetFigFont{12}{14.4}{\rmdefault}{\mddefault}{\updefault}PRDM}}}}}
\put(1362,162){\makebox(0,0)[lb]{\smash{{{\SetFigFont{12}{14.4}{\rmdefault}{\mddefault}{\updefault}RDM}}}}}
\end{picture}
}
\end{center}
\caption{Relationship Among FRDM, PRDM, PIFRDM and RDM}
\label{fig1}
\end{figure}

In this paper, we present a new relational data model -- paraconsistent intuitionistic fuzzy relational data
model (PIFRDM).
Our model is based on the paraconsistent intuitionistic fuzzy set theory 
which is an extension of intuitionistic fuzzy set theory\cite{gb93} and 
is capable of 
manipulating incomplete as well as inconsistent information. We use 
both truth-membership function grade $\alpha$ and false-membership 
function grade $\beta$ 
to denote the
status of a tuple of a certain relation with $\alpha, \beta \in [0,1]$
and $\alpha + \beta \leq 2$.  
PIFRDM is the 
generalization of fuzzy relational data model (FRDM). 
That is, when $\alpha + \beta = 1$, paraconsistent intuitionistic fuzzy 
relation is the ordinary fuzzy relation. This model is distinct with 
paraconsistent relational
data model (PRDM), in fact it can be easily shown that PRDM is
a special case of PIFRDM. That is when $\alpha, \beta = 0 \mbox{ or } 1$, 
paraconsistent intuitionistic fuzzy relation is just paraconsistent relation. 
We can use Figure ~\ref{fig1} to express
the relationship among FRDM, PRDM and PIFRDM. 

We introduce paraconsistent intuitionistic fuzzy relations, which are the 
fundamental mathematical structures
underlying our model. These structures are strictly more general than 
classical
fuzzy relations and intuitionistic fuzzy relations 
(interval-valued fuzzy relations), in that for any fuzzy relation or
intuitionistic fuzzy relation (interval-valued fuzzy relation) there 
is a paraconsistent intuitionistic fuzzy relation with
the same information content, but not {\em vice versa}. The claim is also true 
for the relationship between paraconsistent intuitionistic fuzzy relations and paraconsistent relations. We define algebraic
operators over paraconsistent intuitionistic fuzzy relations that 
extend the standard operators such as
selection, join, union over fuzzy relations.

There are many potential applications of our new data model. 
Here are some examples:
\begin{description}
\item[(a)] Web mining. Essentially the data and documents on the Web are 
heterogeneous, inconsistency is unavoidable. Using the presentation and
reasoning method of our data model, it is easier to capture imperfect 
information on the Web which will provide more potentially value-added 
information. 
\item[(b)] Bioinformatics. There is a proliferation of data sources. Each
research group and each new experimental technique seems to generate yet
another source of valuable data. But these data can be incomplete and 
imprecise and even inconsistent. We could not simply throw away one data 
in favor of
other data. So how to represent and extract useful information
from these data will be a challenge problem. 
\item[(c)] Decision Support System. In decision support system, we need to
combine the database with the knowledge base. There will be a lot of 
uncertain
and inconsistent information, so we need an efficient data model to 
capture
these information and reasoning with these information.
\end{description}

The paper is organized as follow. Section \ref{Fuzzy_Relation} 
of the paper deals with some of the
basic definitions and concepts of fuzzy relations and operations. 
Section \ref{Vague_Relation}  
introduces paraconsistent intuitionistic fuzzy relations and two notions of 
generalising the fuzzy relational
operators such as union, join, projection for these relations. 
Section \ref{Generalized} 
presents some actual generalised algebraic operators for paraconsistent 
intuitionistic fuzzy relations.
These operators can be used for sepcifying queries for database systems 
built
on such relations. 
Section \ref{Application}  gives an illustrative application of these 
operators. 
Finally, Section \ref{Conclusion} contains some concluding remarks and 
directions
for future work. 

\section{Fuzzy Relations and Operations} \label{Fuzzy_Relation}

In this section, we present the essential concepts
of a fuzzy relational database. Fuzzy relations associate 
a value between 0 and 1 with every tuple representing the
degree of membership of the tuple in the relation.
We also present several useful query operators on
fuzzy relations.

Let a {\em relation scheme} (or just {\em scheme})
$\Sigma$ be a finite set of {\em attribute names},
where for any attribute name
$A \in \Sigma$, $\mbox{\em dom}(A)$ is a non-empty
{\em domain} of values for $A$.
A {\em tuple} on $\Sigma$ is any map $t:\Sigma \rightarrow \cup_{A
\in \Sigma}
\mbox{\em dom}(A)$, such that $t(A) \in \mbox{\em dom}(A)$, for each
$A \in \Sigma$.
Let $\tau(\Sigma)$ denote the set of all tuples on $\Sigma$.

\begin{definition}
A {\em fuzzy relation} on scheme $\Sigma$ is any 
map $R: \tau(\Sigma) \rightarrow [0,1]$.
We let ${\cal F}(\Sigma)$ be the set of all fuzzy relations on
$\Sigma$.
$\Box$
\end{definition}                                                                

If $\Sigma$ and $\Delta$ are relation schemes such that $\Delta  
\subseteq \Sigma$, then for any tuple $t \in \tau(\Delta)$, 
we let $t^\Sigma$ denote the set $\{t' \in \tau(\Sigma) ~|~ t'(A) = 
t(A)\mbox{, for all  $A \in \Delta$}\}$ of all extensions of $t$.
We extend this notion for any $T \subseteq \tau(\Delta)$ by defining
$T^\Sigma = \cup_{t \in T} ~ t^\Sigma$.

\subsection{Set-theoretic operations on Fuzzy relations}

\begin{definition}
{\bf Union:} Let $R$ and $S$ be fuzzy relations on scheme $\Sigma$.
Then, $R \cup S$ is a fuzzy relation on scheme $\Sigma$ given by
\[ 
 (R \cup S)(t) =  \max \{ R(t), S(t) \}, \mbox{for any } t \in \tau(\Sigma).\Box
\]

\end{definition}

\begin{definition}
{\bf Complement:} Let $R$ be a fuzzy relation on scheme $\Sigma$.
Then, $-R$ is a fuzzy relation on scheme $\Sigma$ given by
\[
 (-R)(t) = 1 - R(t), \mbox{for any } t \in \tau(\Sigma).\Box
\]
\end{definition}

\begin{definition}
{\bf Intersection:} Let $R$ and $S$ be fuzzy relations on scheme $\Sigma$.
Then, $R \cap S$ is a fuzzy relation on scheme $\Sigma$ given by
\[
 (R \cap S)(t) = \min\{ R(t), S(t) \}, \mbox{for any }t \in \tau(\Sigma).\Box
\]
\end{definition}

\begin{definition}
{\bf Difference:} Let $R$ and $S$ be fuzzy relations on scheme $\Sigma$.
Then, $R - S$ is a fuzzy relation on scheme $\Sigma$ given by
\[
 (R - S)(t) = \min\{ R(t), 1 - S(t) \}, \mbox{for any } t \in \tau(\Sigma).\Box
\]
\end{definition}

\subsection{Relation-theoretic operations on Fuzzy relations}

\begin{definition}
Let $R$ and $S$ be fuzzy relations on schemes $\Sigma$ and  
$\Delta$, respectively.
Then, the {\em natural join} (or just {\em join}) of $R$ and $S$,
denoted $R \Join S$, is a fuzzy relation on scheme  
$\Sigma \cup \Delta$, given by
\[
 (R \Join S)(t) = \min \{ R(\pi_{\Sigma}(t)), S(\pi_{\Delta}(t)) \}, 
\mbox{for any }t \in \tau(\Sigma \cup \Delta).\Box
\]
\end{definition}

\begin{definition}
Let $R$ be a fuzzy relation on scheme $\Sigma$ and let 
$\Delta \subseteq \Sigma$. Then, the projection of $R$ onto
$\Delta$, denoted by $\Pi_{\Delta}(R)$ is a fuzzy relation
on scheme $\Delta$ given by
\[
(\Pi_{\Delta}(R))(t) = \max \{ R(u) | u \in t^{\Sigma} \},
\mbox{for any }t \in \tau(\Delta).\Box
\]
\end{definition}

\begin{definition}
Let $R$ be a fuzzy relation on scheme $\Sigma$,
and let $F$ be any logic formula involving attribute names in  
$\Sigma$,
constant symbols
(denoting values in the attribute domains), equality symbol $=$,
negation symbol $\neg$, and connectives $\vee$ and $\wedge$.
Then, the {\em selection} of $R$ by $F$,
denoted $\dot{\sigma}_{F}(R)$, is a fuzzy relation on scheme  
$\Sigma$,
given by
\[
(\dot{\sigma}_{F}(R))(t) = \left \{ \begin{array}{ll}
R(t) & \mbox{if }t \in \sigma_{F}(\tau(\Sigma)) \\
0 & \mbox{Otherwise}
\end{array} \right. 
\]
where $\sigma_F$ is the usual selection of tuples satisfying $F$.
$\Box$
\end{definition}

\section{Paraconsistent Intuitionistic Fuzzy Relations} 
\label{Vague_Relation}

In this section, we generalize
fuzzy relations in such a manner that we are now able to assign
a measure of belief and a measure of doubt to each tuple. 
We shall refer to these
generalized fuzzy relations as 
{\em paraconsistent intuitionistic fuzzy relations}. 
So, a tuple in a paraconsistent intuitionistic fuzzy relation is 
assigned a measure 
$\langle \alpha, \beta \rangle$, $0 \leq \alpha, \beta \leq 1$.
$\alpha$ will be referred to as the {\em belief} factor and $\beta$ will
be referred to as the {\em doubt} factor.
The interpretation of this measure is that we believe with 
confidence $\alpha$ and doubt with confidence $\beta$ that the tuple is in 
the relation. The belief and doubt confidence factors for a tuple
need not add to exactly 1. This allows for incompleteness and inconsistency
to be represented. If the belief and doubt factors add up to less than 1, 
we have incomplete information regarding the tuple's status in the relation
and if the belief and doubt factors add up to more than 1, we have 
inconsistent
information regarding the tuple's status in the relation.

In contrast to fuzzy relations where the grade of membership of
a tuple is fixed, paraconsistent intuitionistic fuzzy relations bound the grade of membership
of a tuple to a subinterval $[\alpha, 1- \beta]$ for the case 
$\alpha + \beta \leq 1$.

The operators on fuzzy relations can also be generalised for  
paraconsistent intuitionistic fuzzy relations.
However, any such generalisation of operators should maintain the belief  
system
intuition behind paraconsistent intuitionistic fuzzy relations.

This section also develops two different notions of operator
generalisations.

We now formalize the notion of a paraconsistent intuitionistic fuzzy relation.

Recall that $\tau(\Sigma)$ denotes the set of all tuples on any scheme 
$\Sigma$.

\begin{definition}
A {\em paraconsistent intuitionistic fuzzy relation} $R$ on scheme $\Sigma$ is any map 
\[
 R : \tau(\Sigma) \rightarrow [0,1] \times [0,1].
\]
For any $t \in \tau{(\Sigma)}$, we shall denote
$R(t) = \langle R(t)^{+},R(t)^{-} \rangle$, where $R(t)^{+}$ is the belief factor assigned to
$t$ by $R$ and $R(t)^{-}$ is the doubt factor assigned to $t$ by $R$. 
We let ${\cal V}(\Sigma)$ be the set of all paraconsistent intuitionistic fuzzy relations  
on $\Sigma$. $\Box$
\end{definition}

\begin{definition}
A paraconsistent intuitionistic fuzzy relation $R$ on scheme
$\Sigma$ is {\em consistent} if $R(t)^{+} + R(t)^{-} \leq 1$,
for all $t \in \tau(\Sigma)$.
We let ${\cal C}(\Sigma)$ be the set of all consistent paraconsistent intuitionistic fuzzy relations
on $\Sigma$.
Moreover, $R$ is said to be {\em complete} if $R(t)^{+} + R(t)^{-} \geq 1$,
for all $t \in \tau(\Sigma)$.  
If $R$ is both consistent and complete,
i.e. $R(t)^{+} + R(t)^{-} = 1$,
for all $t \in \tau(\Sigma)$,
then it is a {\em total paraconsistent intuitionistic fuzzy relation},
and we let ${\cal T}(\Sigma)$ be the set of all total paraconsistent intuitionistic fuzzy
relations on $\Sigma$.
$R$ is said to be {\em pseudo-consistent} if
$\max(R(t_i)^+) + \max(R(t_i)^-) > 1, R(t_i)^+ + R(t_i)^- = 1$,
for some $t_i \in \tau(\Sigma)$, these $t_i$'s have the same values on
$\Sigma$ with different belief factor and doubt factor and for all the other
$t \in tau(\Sigma), R(t)^+ + R(t)^- \leq 1$.
We let ${\cal P}(\Sigma)$ be the set of all pseudo-consistent paraconsistent
intuitionistic fuzzy relations on $\Sigma$.
$\Box$
\end{definition}

Note that pseudo-consistent paraconsistent intuitionistic fuzzy relation is
a subclass of consistent paraconsistent intuitionistic fuzzy relation.

It should be observed that total paraconsistent intuitionistic fuzzy relations
are essentially fuzzy relations where the uncertainity in the
grade of membership is eliminated.
We make this relationship explicit by defining a one-one  
correspondence
$\lambda_\Sigma:{\cal T}(\Sigma) \rightarrow {\cal F}(\Sigma)$, given  
by
$\lambda_\Sigma(R)(t) = R(t)^{+}$, for all $t \in \tau(\Sigma)$.
This correspondence is used frequently in the following discussion.

\subsection*{Operator Generalisations}

It is easily seen that paraconsistent intuitionistic fuzzy relations are a 
generalisation  of
fuzzy relations, in that for each fuzzy relation there is a
paraconsistent intuitionistic fuzzy relation with the same information content, but not  
{\em vice versa}.
It is thus natural to think of generalising the operations on  
fuzzy relations such as union, join, projection etc. to paraconsistent 
intuitionistic fuzzy  
relations.
However, any such generalisation should be intuitive with respect to
the belief system model of paraconsistent intuitionistic fuzzy relations.
We now construct a framework for operators on both kinds of relations
and introduce two different notions of the generalisation  
relationship among their operators.

An $n$-ary {\em operator on fuzzy relations with signature
$\langle \Sigma_1,\ldots,\Sigma_{n+1} \rangle$} is a function
$\Theta:{\cal F}(\Sigma_1) \times \cdots \times {\cal F}(\Sigma_n)
\rightarrow {\cal F}(\Sigma_{n+1})$, where
$\Sigma_1,\ldots,\Sigma_{n+1}$ are any schemes.
Similarly, an $n$-ary {\em operator on paraconsistent intuitionistic 
fuzzy relations with  
signature
$\langle \Sigma_1,\ldots,\Sigma_{n+1} \rangle$} is a function
$\Psi:{\cal V}(\Sigma_1) \times \cdots \times {\cal V}(\Sigma_n)
\rightarrow {\cal V}(\Sigma_{n+1})$.

\begin{definition}
An operator $\Psi$ on paraconsistent intuitionistic fuzzy relations
with signature $\langle \Sigma_1,\ldots,\Sigma_{n+1} \rangle$
is {\em totality preserving} if for any
total paraconsistent intuitionistic fuzzy relations $R_1,\ldots,R_n$ on 
schemes
$\Sigma_1,\ldots,\Sigma_n$, respectively, $\Psi(R_1,\ldots,R_n)$
is also total.
$\Box$
\end{definition}

\begin{definition}
A totality preserving operator
$\Psi$ on paraconsistent intuitionistic fuzzy relations
with signature 
\begin{center}
$\langle \Sigma_1,\ldots,\Sigma_{n+1} \rangle$
\end{center}
is a {\em weak generalisation} of an
operator
$\Theta$ on fuzzy relations with the same signature, if for
any total paraconsistent intuitionistic fuzzy relations $R_1,\ldots,R_n$ on 
schemes
$\Sigma_1,\ldots,\Sigma_n$, respectively, we have
\[
 \lambda_{\Sigma_{n+1}}(\Psi(R_1,\ldots,R_n)) =
\Theta(\lambda_{\Sigma_1}(R_1),\ldots,\lambda_{\Sigma_n}(R_n)). 
\Box
\]
\end{definition}
The above definition essentially requires $\Psi$ to coincide with
$\Theta$ on total paraconsistent intuitionistic fuzzy relations 
(which are in one-one
correspondence with the fuzzy relations).
In general, there may be many operators on paraconsistent intuitionistic 
fuzzy relations
that are weak generalisations
of a given operator $\Theta$ on fuzzy relations.
The behavior of the weak generalisations of $\Theta$ on even just the
consistent paraconsistent intuitionistic fuzzy relations may in general vary.
We require a stronger notion of operator generalisation under which,  
at
least when restricted to consistent 
intuitionistic fuzzy 
relations, the behavior of all the generalised operators is the
same.
Before we can develop such a notion, we need that of `representations' of  
a
paraconsistent intuitionistic fuzzy relation.

We associate with a consistent paraconsistent 
intuitionistic fuzzy relation $R$ the set of
all (fuzzy relations corresponding to) total paraconsistent intuitionistic 
fuzzy relations
obtainable from $R$ by filling in the gaps between the belief and
doubt factors for each tuple.
Let the map $\mbox{\bf reps}_\Sigma:{\cal C}(\Sigma) \rightarrow  
2^{{\cal F}
(\Sigma)}$ be given by
\[
\mbox{\bf reps}_\Sigma(R) = \{Q \in {\cal F}(\Sigma) ~|~ 
\bigwedge_{t_{i} \in \tau(\Sigma)}  
(R(t_{i})^{+} \leq Q(t_{i}) \leq 1 - R(t_{i})^{-} ) \}.
\]
The set $\mbox{\bf reps}_\Sigma(R)$ contains all fuzzy relations  
that are `completions' of the consistent paraconsistent 
intuitionistic fuzzy relation $R$.
Observe that $\mbox{\bf reps}_\Sigma$ is defined only for consistent  
paraconsistent intuitionistic fuzzy relations and produces sets of 
fuzzy relations. Then we have following 
observation.

\begin{proposition} \label{prop1}
 For any consistent paraconsistent intuitionistic fuzzy relation $R$ on 
 scheme $\Sigma$, $\mbox{\bf reps}_\Sigma(R)$ is the singleton 
 $\{\lambda_\Sigma(R)\} \mbox{ iff }R \mbox{ is total}.\Box$
\end{proposition}
\begin{proof}
It is clear from the definition of consistent and total paraconsistent
intuitionistic fuzzy relations and from the definition of {\bf reps}
operation.
\end{proof}

We now need to extend operators on fuzzy relations to sets of
fuzzy relations.
For any operator
$\Theta:{\cal F}(\Sigma_1) \times \cdots \times {\cal F}(\Sigma_n)
\rightarrow {\cal F}(\Sigma_{n+1})$
on fuzzy relations, we let ${\cal S}(\Theta):
2^{{\cal F}(\Sigma_1)} \times \cdots \times 2^{{\cal F}(\Sigma_n)}
\rightarrow 2^{{\cal F}(\Sigma_{n+1})}$ be a map on
sets of fuzzy relations defined as follows.
For any sets $M_1,\ldots,M_n$ of fuzzy relations on schemes
$\Sigma_1,\ldots,\Sigma_n$, respectively,
\[
 {\cal S}(\Theta)(M_1,\ldots,M_n) = \{\Theta(R_1,\ldots,R_n) ~|~ R_i  
\in M_i,
\mbox{ for all } i, 1\leq i\leq n\}.
\]
In other words, ${\cal S}(\Theta)(M_1,\ldots,M_n)$ is the set of
$\Theta$-images of all tuples in the cartesian product
$M_1 \times \cdots \times M_n$.
We are now ready to lead up to a stronger notion of operator
generalisation.

\begin{definition}
An operator $\Psi$ on paraconsistent intuitionistic fuzzy relations
with signature $\langle \Sigma_1,\ldots,\Sigma_{n+1} \rangle$
is {\em consistency preserving} if for any
consistent paraconsistent intuitionistic fuzzy relations $R_1,\ldots,R_n$ on schemes
$\Sigma_1,\ldots,\Sigma_n$, respectively, $\Psi(R_1,\ldots,R_n)$
is also consistent.
$\Box$
\end{definition}

\begin{definition}
A consistency preserving operator
$\Psi$ on paraconsistent intuitionistic fuzzy relations
with signature $\langle \Sigma_1,\ldots,\Sigma_{n+1} \rangle$
is a {\em strong generalisation} of an operator
$\Theta$ on fuzzy relations with the same signature, if
for any consistent paraconsistent intuitionistic fuzzy relations $R_1,\ldots,R_n$ on schemes
$\Sigma_1,\ldots,\Sigma_n$, respectively,
we have
\[
 \mbox{\bf reps}_{\Sigma_{n+1}}(\Psi(R_1,\ldots,R_n)) =
{\cal S}(\Theta)(\mbox{\bf reps}_{\Sigma_1}(R_1),\ldots,
\mbox{\bf reps}_{\Sigma_n}(R_n)). \Box
\]
\end{definition}

Given an operator $\Theta$ on fuzzy relations, the behavior of a
weak generalisation of $\Theta$ is `controlled' only over the total
paraconsistent intuitionistic fuzzy relations.
On the other hand, the behavior of a strong generalisation is
`controlled' over all consistent paraconsistent intuitionistic fuzzy relations.
This itself suggests that strong generalisation is a stronger notion
than weak generalisation.
The following proposition makes this precise.

\begin{proposition}
If $\Psi$ is a strong generalisation of $\Theta$,
then $\Psi$ is also a weak generalisation of $\Theta$.$\Box$
\end{proposition}
\begin{proof}
Let $\langle \Sigma_1, \ldots, \Sigma_{n+1} \rangle$ be the signature
of $\Psi$ and $\Theta$, and let $R_1, \ldots, R_n$ be any total paraconsistent intuitionistic fuzzy
relations on schemes $\Sigma_1, \ldots, \Sigma_n$, respectively. Since
all total relations are consistent, and $\Psi$ is a strong generalisation
of $\Theta$, we have that
\[
 {\bf reps}_{\Sigma_{n+1}}(\Psi(R_1, \ldots, R_n)) = {\cal S}(\Theta)({\bf reps}_{\Sigma_1}(R_1), \ldots, {\bf reps}_{\Sigma_n}(R_n)),
\]
Proposition~\ref{prop1} gives us that for each $i$, $1 \leq i \leq n$,
${\bf reps}_{\Sigma_i}(R_i)$ is the singleton set $\{\lambda_{\Sigma_i}(R_i)\}$.
Therefore, 
${\cal S}(\Theta)({\bf reps}_{\Sigma_1}(R_i), \ldots, {\bf reps}_{\Sigma_n}(R_n))$ 
is just the singleton set:
\begin{center}
$\{\Theta(\lambda_{\Sigma_1}(R_1), \ldots, \lambda_{\Sigma_n}(R_n))\}$. 
\end{center}
Here, $\Psi(R_1, \ldots, R_n)$ is total, and \\
$\lambda_{\Sigma_{n+1}}(\Psi(R_1, \ldots, R_n)) = \Theta(\lambda_{\Sigma_1}(R_1), \ldots, \lambda_{\Sigma_n}(R_n))$, i.e. $\Psi$ is a weak generalisation of $\Theta$.
\end{proof}

Though there may be many strong generalisations of an operator on  
fuzzy
relations, they all behave the same when restricted to consistent
paraconsistent intuitionistic fuzzy relations.
In the next section, we propose strong generalisations for the usual
operators on fuzzy relations.
The proposed generalised operators on paraconsistent intuitionistic fuzzy  relations  
correspond to the
belief system intuition behind paraconsistent intuitionistic fuzzy  relations.

First we will introduce two special operators on paraconsistent
intuitionistic fuzzy relations called split and combine to
transform inconsistent paraconsistent intuitionistic fuzzy
relations into pseudo-consistent paraconsistent intuitionistic
fuzzy relations and transform pseudo-consistent paraconsistent
intuitionistic fuzzy relations into inconsistent paraconsistent
intuitionistic fuzzy relations.

\begin{definition}({\bf Split})
Let $R$ be paraconsistent intuitionistic fuzzy
relations on scheme $\Sigma$.
Then,
$\triangle(R) = \{t_i \in R | (R(t_i)^+ + R(t_i)^- > 1)  \wedge (\triangle(R)$
$(t_i)^+ = R(t_i)^+ \wedge \triangle(R)(t_i)^- = 1 - R(t_i)^+ \vee \triangle(R)$
$(t_i)^+ = 1 - R(t_i)^- \wedge \triangle(R)(t_i)^- = R(t_i)^-) \vee $
$(R(t_i)^+ +$ 
$R(t_i)^- \leq 1) \wedge (\triangle(R)(t_i)^+ = R(t_i)^+ \wedge \triangle(R)(t_i)^- = R(t_i)^-)\}$.
$\Box$
\end{definition}

It is obvious that $\triangle(R)$ is
pseudo-consistent if $R$ is inconsistent.

\begin{definition}({\bf Combine})
Let $R$ be paraconsistent intuitionistic fuzzy relations
on scheme $\Sigma$.
Then,
$\nabla(R) = \{t_i \in R | (\forall i)(\nabla(R)(t_i)^+ =\max(R(t_i)^+) \wedge 
(\nabla(R)(t_i)^- = \max(R(t_i)^-) \}$
\end{definition}

It is obvious that $\nabla(R)$ is inconsistent if $R$ is pseudo-consistent.

Note that strong generalization defined above only holds for consistent or
pseudo-consistent paraconsistent intuitionistic fuzzy relations. For any
arbitrary paraconsisent intuitionistic fuzzy relations, we should first
use split operation to transform them into non inconsistent paraconsistent
intuitionistic fuzzy relations and apply the set-theoretic and 
relation-theoretic
operations on them and finally use combine operation to transform the result
into arbitrary paraconsistent intuitionistic fuzzy relation. For the simplificat
ion
of notation, the following generalized algebra is defined under such assumption.

\section{Generalized Algebra on Paraconsistent Intuitionistic Fuzzy Relations} 
\label{Generalized}

In this section, we present one strong generalisation each for the  
fuzzy relation operators such as union, join, projection.
To reflect generalisation, a hat is placed over
a fuzzy relation operator to
obtain the corresponding paraconsistent intuitionistic fuzzy relation operator.
For example, $\bowtie$ denotes the natural join among fuzzy  
relations, and
$\widehat{\bowtie}$ denotes natural join on paraconsistent intuitionistic fuzzy relations.
These generalized operators maintain the belief system
intuition behind paraconsistent intuitionistic fuzzy relations.

\subsection*{Set-Theoretic Operators}

We first generalize the two fundamental set-theoretic operators,
union and complement.

\begin{definition}
Let $R$ and $S$ be paraconsistent intuitionistic fuzzy relations on scheme $\Sigma$.
Then,
\begin{description}
\item[(a)] the {\em union} of $R$ and $S$, denoted $R~\widehat{\cup}~S$,  
is a paraconsistent intuitionistic fuzzy relation on scheme $\Sigma$, given
by
\[
(R ~\widehat{\cup}~ S)(t) = \langle \max\{R(t)^{+},S(t)^{+}\}, \min\{R(t)^{-},S(t)^{-}\} \rangle,
\mbox{ for any } t \in \tau(\Sigma);
\]
\item[(b)] the {\em complement} of $R$, denoted $\widehat{-}~R$, is a
paraconsistent intuitionistic fuzzy relation on scheme $\Sigma$, given
by
\[
(\widehat{-}~R)(t) = \langle R(t)^{-},R(t)^{+} \rangle, \mbox{ for any } t \in \tau(\Sigma).
~
\]
\hfill{\space}  $\Box$
\end{description}
\end{definition}
An intuitive appreciation of the union operator can be obtained as follows:
Given a tuple $t$, since we believed that it is present in the relation $R$
with confidence $R(t)^{+}$ and that it is present in the relation $S$
with confidence $S(t)^{+}$, we can now believe that the tuple $t$ is
present in the ``either-$R$-or-$S$'' relation with confidence which is equal to
the larger of $R(t)^{+}$ and $S(t)^{+}$. Using the same logic, we can now
believe in the absence of the tuple $t$ from the ``either-$R$-or-$S$'' relation
with confidence which is equal to the smaller (because $t$ must
be absent from both $R$ and $S$ for it to be absent from the union) of
$R(t)^{-}$ and $S(t)^{-}$. The definition of {\em complement} and
of all the other operators on paraconsistent intuitionistic fuzzy relations defined
later can (and should) be understood in the same way.

\begin{proposition}
The operators $\widehat{\cup}$ and unary $\widehat{-}$ on paraconsistent intuitionistic fuzzy  
relations are
strong generalisations of the operators $\cup$ and unary $-$ on
fuzzy relations.
\end{proposition}
\begin{proof}
Let $R$ and $S$ be consistent paraconsistent intuitionistic fuzzy relations on scheme $\Sigma$. Then
${\bf reps}_{\Sigma}(R~\widehat{\cup}~S)$ is the set
\[
 \{Q ~|~ \bigwedge_{t_i \in \tau(\Sigma)}(\max\{R(t_i)^+,~S(t_i)^+\} \leq Q(t_i) \leq 1 - \min\{R(t_i)^-,~S(t_i)^-\})\} 
\]
This set is the same as the set 
\[
 \{r~\cup~s ~|~ \bigwedge_{t_i \in \tau(\Sigma)}(R(t_i)^+ \leq r(t_i) \leq 1-R(t_i)^-), \bigwedge_{t_i \in \tau(\Sigma)}(S(t_i)^+ \leq s(t_i) \leq 1-S(t_i)^-)\}
\]
which is $S(\cup)({\bf reps}_{\Sigma}(R),~{\bf reps}_{\Sigma}(S))$. 
Such a result for unary $\widehat{-}$ can also be shown similarly.
\end{proof}

For sake of completeness, we define the following two related  
set-theoretic operators:

\begin{definition}
Let $R$ and $S$ be paraconsistent intuitionistic fuzzy relations on scheme $\Sigma$.
Then,
\begin{description}
\item[(a)] the {\em intersection} of $R$ and $S$, denoted  
$R~\widehat{\cap}~S$,
is a paraconsistent intuitionistic fuzzy relation on scheme $\Sigma$, given
by
\[
(R ~\widehat{\cap}~ S)(t) = \langle \min\{R(t)^{+},S(t)^{+}\}, \max\{R(t)^{-},S(t)^{-}\} \rangle,
\mbox{ for any } t \in \tau(\Sigma);
\]
\item[(b)] the {\em difference} of $R$ and $S$, denoted  
$R~\widehat{-}~S$, is
a paraconsistent intuitionistic fuzzy relation on scheme $\Sigma$, given
by
\[
(R~\widehat{-}~S)(t) = \langle \min\{R(t)^{+},S(t)^{-}\}, \max\{R(t)^{-},S(t)^{+}\} \rangle,
\mbox{ for any } t \in \tau(\Sigma);
\]
\hfill{\space}  $\Box$
\end{description}
\end{definition}

The following proposition relates the intersection and difference
operators in terms of the more fundamental set-theoretic operators
union and complement.

\begin{proposition}
For any paraconsistent intuitionistic fuzzy relations $R$ and $S$ on the same scheme, 
we have
\begin{eqnarray*}
R~\widehat{\cap}~S &=& \widehat{-}(\widehat{-}R~\widehat{\cup}~\widehat{-}S), \mbox{  and  
}\\
R~\widehat{-}~S &=& \widehat{-}(\widehat{-}R~\widehat{\cup}~S).
\end{eqnarray*}
\end{proposition}
\begin{proof}
\begin{eqnarray*}
\mbox{By definiton, } \widehat{-}R(t) &=& \langle R(t)^{-}, R(t)^{+} \rangle \\
                     \widehat{-}S(t) &=& \langle S(t)^{-}, S(t)^{+} \rangle \\
\mbox{ and }       (\widehat{-}R~\widehat{\cup}~\widehat{-}S)(t) &=& \langle \max(R(t)^{-}, S(t)^{-}),~\min(R(t)^{+}, S(t)^{+}) \rangle \\
\mbox{ so, }    (\widehat{-}(\widehat{-}R~\widehat{\cup}~\widehat{-}S))(t) &=& \langle \min(R(t)^{+}, S(t)^{+}), \max(R(t)^{-}, S(t)^{-}) \rangle \\
                    &=& R~\widehat{\cap}~S(t).
\end{eqnarray*}
The second part of the result can be shown similarly.
\end{proof}

\subsection*{Relation-Theoretic Operators}

We now define some relation-theoretic algebraic
operators on paraconsistent intuitionistic fuzzy relations.

\begin{definition}
Let $R$ and $S$ be paraconsistent intuitionistic fuzzy relations on 
schemes $\Sigma$ and  $\Delta$, respectively.
Then, the {\em natural join} (further for short called {\em join}) 
of $R$ and $S$,
denoted $R~\widehat{\bowtie}~S$, is a paraconsistent intuitionistic fuzzy relation on scheme  $\Sigma \cup \Delta$,
given by
\[
(R~\widehat{\bowtie}~S)(t) = 
\langle \min \{R(\pi_{\Sigma}(t))^{+}, S(\pi_{\Delta}(t))^{+} \}, 
 \max \{R(\pi_{\Sigma}(t))^{-}, S(\pi_{\Delta}(t))^{-} \} \rangle,
\]
where $\pi$ is the usual projection of a tuple.
\hfill{\space} $\Box$
\end{definition}
It is instructive to observe that, similar to the intersection operator,
the minimum of the belief factors and the maximum of the doubt factors 
are used in the definition of the join operation. 

\begin{proposition} \label{strjoin}
$\widehat{\bowtie}$ is a strong generalisation of $\bowtie$.
\end{proposition}
\begin{proof}
Let $R$ and $S$ be consistent paraconsistent intuitionistic fuzzy relations on schemes $\Sigma$ and $\Delta$,
respectively. Then ${\bf reps}_{\Sigma~\cup~\Delta}(R~\widehat{\bowtie}~S)$ is 
the set
$\{Q \in {\cal F}(\Sigma~\cup~\Delta)~|~\bigwedge_{t_i \in \tau(\Sigma~\cup~\Delta)}(\min \{R_{\pi_\Sigma}(t_i)^+, S_{\pi_\Delta}(t_i)^+ \} \leq Q(t_i) \leq$ \\
$1-\max \{R_{\pi_\Sigma}(t_i)^-, S_{\pi_\Delta}(t_i)^- \}) \}$
and $S(\bowtie)({\bf reps}_\Sigma(R), {\bf reps}_\Delta(S)) = \{ r \bowtie s ~|~r \in {\bf reps}_\Sigma(R), s \in {\bf reps}_\Delta(S) \}$

Let $Q \in {\bf reps}_{\Sigma \cup \Delta}(R~\widehat{\bowtie}~S)$. Then
$\pi_\Sigma(Q)\in {\bf reps}_\Sigma(R)$, where $\pi_\Sigma$ is the usual
projection over $\Sigma$ of fuzzy relations. 
Similarly, $\pi_\Delta(Q) \in {\bf reps}_\Delta(S)$.  
Therefore, $Q \in S(\bowtie)({\bf reps}_\Sigma(R), {\bf reps}_\Delta(S))$.

Let $Q \in S(\bowtie)({\bf reps}_\Sigma(R), {\bf reps}_\Delta(S))$.
Then $Q(t_i) \geq \min \{R_{\pi_\Sigma}(t_i)^+, S_{\pi_\Delta}(t_i)^+ \}$
and $Q(t_i) \leq \min \{1-R{\pi_\Sigma}(t_i)^-, 1-S_{\pi_\Delta}(t_i)^- \} = 1-\max \{R_{\pi_\Sigma}(t_i)^-, S_{\pi_\Delta}(t_i)^- \}$, for any $t_i \in \tau(\Sigma \cup \Delta)$, because $R$ and $S$ are consistent. \\
Therefore, $Q \in {\bf reps}_{\Sigma \cup \Delta}(R~\widehat{\bowtie}~S)$.
\end{proof}

We now present the projection operator.

\begin{definition}
Let $R$ be a paraconsistent intuitionistic fuzzy relation on scheme $\Sigma$, and
$\Delta \subseteq \Sigma$.
Then, the {\em projection} of $R$ onto $\Delta$,
denoted $\widehat{\pi}_\Delta(R)$, is a paraconsistent intuitionistic fuzzy relation on  
scheme $\Delta$, given by
\[
(\widehat{\pi}_\Delta(R))(t) = 
\langle \max\{R(u)^{+} | u \in t^{\Sigma} \},
 \min\{R(u)^{-} | u \in t^{\Sigma} \}  \rangle.
\]
\hfill{\space} $\Box$
\end{definition}
The belief factor of a tuple in the projection is the maximum
of the belief factors of all of the tuple's extensions onto the
scheme of the input paraconsistent intuitionistic fuzzy relation.
Moreover, the doubt factor of a tuple in the projection is the minimum
of the doubt factors of all of the tuple's extensions onto the
scheme of the input paraconsistent intuitionistic fuzzy relation.

We present the selection operator next.

\begin{definition}
Let $R$ be a paraconsistent intuitionistic fuzzy relation on scheme $\Sigma$,
and let $F$ be any logic formula involving attribute names in  
$\Sigma$, constant symbols
(denoting values in the attribute domains), equality symbol $=$,
negation symbol $\neg$, and connectives $\vee$ and $\wedge$.
Then, the {\em selection} of $R$ by $F$,
denoted $\widehat{\sigma}_F(R)$, is a paraconsistent intuitionistic fuzzy relation on
scheme  $\Sigma$, given by

\vspace{.15in}

\begin{tabular}{l}
$(\widehat{\sigma}_F(R))(t) = \langle \alpha,\beta \rangle$, where \\[.15in]
$\alpha = \left\{ \begin{array}{ll}
                    R(t)^{+} & \mbox{ if $t \in \sigma_{F}(\tau(\Sigma))$} \\
                    0        & \mbox{otherwise}
                   \end{array}
           \right. $  \hspace{0.25in} and \hspace{0.25in}
$\beta  = \left\{ \begin{array}{ll}
                    R(t)^{-} & \mbox{ if $t \in \sigma_{F}(\tau(\Sigma))$} \\
                    1        & \mbox{otherwise}
                   \end{array}
           \right. $
\end{tabular} \\[.15in] 
where $\sigma_F$ is the usual selection of tuples satisfying $F$ from
ordinary relations.
\hfill{\space}$\Box$
\end{definition}
If a tuple satisfies the selection criterion, it's belief and doubt factors
are the same in the selection as in the input paraconsistent intuitionistic fuzzy
relation. In the case where the tuple does not satisfy the selection 
criterion, its belief factor is set to 0 and the doubt factor is set
to 1 in the selection.

\begin{proposition}
The operators $\widehat{\pi}$ and $\widehat{\sigma}$ are strong  
generalisations of
$\pi$ and $\sigma$, respectively.
\end{proposition}
\begin{proof}
Similar to that of Proposition~\ref{strjoin}.
\end{proof}

\begin{example}
Relation schemes are sets of attribute names, but in this example we treat 
them as ordered sequences of attribute names (which can be obtained through
permutation of attribute names), so tuples can be viewed as the usual lists 
of values. Let $\{a, b, c\}$ be a common domain for all attribute names, and 
let $R$ and $S$ be the following
paraconsistent intuitionistic fuzzy relations on schemes $\langle X,Y \rangle$ and $\langle Y,Z \rangle$
respectively.
\begin{center}
\begin{tabular}{|c|c|} \hline
$t$        & $R(t)$ \\ \hline
$(a,a)$    & $\langle 0,1 \rangle$ \\ 
$(a,b)$    & $\langle 0,1 \rangle$ \\
$(a,c)$    & $\langle 0,1 \rangle$ \\
$(b,b)$    & $\langle 1,0 \rangle$ \\
$(b,c)$    & $\langle 1,0 \rangle$ \\ 
$(c,b)$    & $\langle 1,1 \rangle$ \\ \hline 
\end{tabular}
\hspace{0.5in}
\begin{tabular}{|c|c|} \hline
$t$        & $S(t)$ \\ \hline
$(a,c)$    & $\langle 1,0 \rangle$ \\
$(b,a)$    & $\langle 1,1 \rangle$ \\
$(c,b)$    & $\langle 0,1 \rangle$ \\ \hline
\end{tabular}
\end{center}
For other tuples which are not in the paraconsistent intuitionistic fuzzy relations $R(t)$ and $S(t)$, their 
$\langle \alpha,\beta \rangle = \langle 0,0 \rangle$ which means no any 
information available. Because $R$ and $S$ are inconsistent, we first
use split operation to trasform them into pseudo-consistent and apply the
relation-theoretic operations on them and transform the result back to
arbitrary paraconsistent intuitionistic fuzzy set using combine operation.
Then, $T_{1} =\nabla(\triangle(R)~\widehat{\bowtie}~\triangle(S))$ is a 
paraconsistent intuitionistic fuzzy 
relation on scheme $\langle X,Y,Z \rangle$ and 
$T_{2} = \nabla(\widehat{\pi}_{\langle X,Z \rangle}(\triangle(T_{1})))$  and 
$T_{3} = \widehat{\sigma}_{X \neg = Z}(T_{2})$ 
are paraconsistent intuitionistic fuzzy relations on scheme 
$\langle X,Z \rangle$. $T_{1}$, $T_{2}$ and $T_{3}$ are 
shown below:
\begin{center}
\begin{tabular}{|c|c|} \hline
$t$       & $T_{1}(t)$ \\ \hline
$(a,a,a)$ & $\langle 0,1 \rangle$    \\
$(a,a,b)$ & $\langle 0,1 \rangle$    \\
$(a,a,c)$ & $\langle 0,1 \rangle$    \\
$(a,b,a)$ & $\langle 0,1 \rangle$    \\
$(a,b,b)$ & $\langle 0,1 \rangle$    \\
$(a,b,c)$ & $\langle 0,1 \rangle$    \\
$(a,c,a)$ & $\langle 0,1 \rangle$    \\
$(a,c,b)$ & $\langle 0,1 \rangle$    \\
$(a,c,c)$ & $\langle 0,1 \rangle$    \\ 
$(b,b,a)$ & $\langle 1,1 \rangle$    \\
$(b,c,b)$ & $\langle 0,1 \rangle$    \\
$(c,b,a)$ & $\langle 1,1 \rangle$    \\
$(c,b,b)$ & $\langle 0,1 \rangle$    \\
$(c,b,c)$ & $\langle 0,1 \rangle$    \\
$(c,c,b)$ & $\langle 0,1 \rangle$    \\ \hline
\end{tabular}
\hspace{.4in}
\begin{tabular}{|c|c|} \hline
$t$       &   $T_2(t)$ \\ \hline
$(a,a)$   &   $\langle 0,1 \rangle$ \\
$(a,b)$   &   $\langle 0,1 \rangle$ \\ 
$(a,c)$   &   $\langle 0,1 \rangle$ \\
$(b,a)$   &   $\langle 1,0 \rangle$ \\ 
$(c,a)$   &   $\langle 1,0 \rangle$ \\ \hline
\end{tabular}
\hspace{0.4in}
\begin{tabular}{|c|c|} \hline
$t$       &   $T_3(t)$ \\ \hline
$(a,a)$   &   $\langle 0,1 \rangle$ \\
$(a,b)$   &   $\langle 0,1 \rangle$ \\
$(a,c)$   &   $\langle 0,1 \rangle$ \\
$(b,a)$   &   $\langle 1,0 \rangle$ \\
$(b,b)$   &   $\langle 0,1 \rangle$ \\
$(c,a)$   &   $\langle 1,0 \rangle$ \\
$(c,c)$   &   $\langle 0,1 \rangle$ \\ \hline
\end{tabular}
\end{center}
\hfill{\space} $\Box$
\end{example}

\section{An Application} \label{Application}

Consider the target recognition example
presented in \cite{sbr94}. Here,
an autonomous vehicle needs to identify
objects in a hostile environment such as a military battlefield.
The autonomous vehicle is equipped with a number of sensors which are used
to collect data, such as speed and size of
the objects (tanks) in the battlefield. 
Associated with each sensor, we have a set of rules that describe the type
of the object based on the properties detected by the sensor.

Let us assume that the autonomous vehicle is equipped with three sensors
resulting in data collected about radar readings, 
of the tanks, their gun characteristics and their speeds.
What follows is a set of rules that associate the type of object with
various observations.

\vspace*{.15in}

\noindent
{\bf Radar Readings:} 
\begin{itemize}
\item Reading $r_1$ indicates that the object is a T-72 tank
with belief factor 0.80 and doubt factor 0.15.
\item Reading $r_2$ indicates that the object is a T-60 tank
with belief factor 0.70 and doubt factor 0.20.
\item Reading $r_3$ indicates that the object is not a T-72 tank
with belief factor 0.95 and doubt factor 0.05.
\item Reading $r_4$ indicates that the object is a T-80 tank
with belief factor 0.85 and doubt factor 0.10.
\end{itemize}

\noindent
{\bf Gun Characteristics:}
\begin{itemize}
\item Characteristic $c_1$ indicates that the object is a T-60 tank
with belief factor 0.80 and doubt factor 0.20.
\item Characteristic $c_2$ indicates that the object is not
a T-80 tank with belief factor 0.90 and doubt factor 0.05.
\item Characteristic $c_3$ indicates that the object is a T-72 tank
with belief factor 0.85 and doubt factor 0.10.
\end{itemize}

\noindent
{\bf Speed Characteristics:}
\begin{itemize}
\item Low speed indicates that the object is a T-60 tank
with belief factor 0.80 and doubt factor 0.15.
\item High speed indicates that the object is not a T-72 tank
with belief factor 0.85 and doubt factor 0.15.
\item High speed indicates that the object is not a T-80 tank
with belief factor 0.95 and doubt factor 0.05.
\item Medium speed indicates that the object is not a T-80 tank
with belief factor 0.80 and doubt factor 0.10.
\end{itemize}

\noindent
These rules can be captured in the following three paraconsistent
intuitionisitic fuzzy relations:

\begin{center}
\begin{tabular}{|c|c||c|}
\multicolumn{3}{c}{RadarRules} \\ \hline
Reading   & Object   & Confidence Factors \\ \hline \hline
$r_1$  & T-72 & $\langle 0.80,0.15 \rangle$ \\ \hline
$r_2$  & T-60 & $\langle 0.70,0.20 \rangle$ \\ \hline
$r_3$  & T-72 & $\langle 0.05,0.95 \rangle$ \\ \hline
$r_4$  & T-80 & $\langle 0.85,0.10 \rangle$ \\ \hline
\end{tabular} \\[0.25in]
\begin{tabular}{|c|c||c|}
\multicolumn{3}{c}{GunRules} \\ \hline
Reading   & Object   & Confidence Factors \\ \hline \hline
$c_1$  & T-60 & $\langle 0.80,0.20 \rangle$ \\ \hline
$c_2$  & T-80 & $\langle 0.05,0.90 \rangle$ \\ \hline
$c_3$  & T-72 & $\langle 0.85,0.10 \rangle$ \\ \hline
\end{tabular} \\[0.25in]
\begin{tabular}{|c|c||c|}
\multicolumn{3}{c}{SpeedRules} \\ \hline
Reading   & Object   & Confidence Factors \\ \hline \hline
low     & T-60 & $\langle 0.80,0.15 \rangle$ \\ \hline
high    & T-72 & $\langle 0.15,0.85 \rangle$ \\ \hline
high    & T-80 & $\langle 0.05,0.95 \rangle$ \\ \hline
medium  & T-80 & $\langle 0.10,0.80 \rangle$ \\ \hline
\end{tabular}
\end{center}

The autonomous vehicle uses the sensors to make observations
about the different objects and then uses the rules
to determine the type of each object in the battlefield.
It is quite possible that two different sensors may identify the same
object as of different types, thereby introducing inconsistencies.

Let us now consider three objects $o_1$, $o_2$ and $o_3$ which need to be identified by the autonomous vehicle. Let us assume the following
observations made by the three sensors about the three objects.
Once again, we assume certainty factors (maybe derived from the accuracy of the sensors) are associated with each observation.

\begin{center}
\begin{tabular}{|c|c||c|} 
\multicolumn{3}{c}{RadarData} \\ \hline
Object-id & Reading & Confidence Factors \\ \hline \hline
$o_1$ & $r_3$ & $\langle 1.00,0.00 \rangle$ \\ \hline
$o_2$ & $r_1$ & $\langle 1.00,0.00 \rangle$ \\ \hline
$o_3$ & $r_4$ & $\langle 1.00,0.00 \rangle$ \\ \hline
\end{tabular} \\[0.25in]
%\hspace{.1in}
\begin{tabular}{|c|c||c|} 
\multicolumn{3}{c}{GunData} \\ \hline
Object-id & Reading & Confidence Factors \\ \hline \hline
$o_1$ & $c_3$ & $\langle 0.80,0.10 \rangle$ \\ \hline
$o_2$ & $c_1$ & $\langle 0.90,0.10 \rangle$ \\ \hline
$o_3$ & $c_2$ & $\langle 0.90,0.10 \rangle$ \\ \hline
\end{tabular} \\[.25in]
%\hspace{.25in}
\begin{tabular}{|c|c||c|} 
\multicolumn{3}{c}{SpeedData} \\ \hline
Object-id & Reading & Confidence Factors \\ \hline \hline
$o_1$ & high   & $\langle 0.90,0.10 \rangle$ \\ \hline
$o_2$ & low    & $\langle 0.95,0.05 \rangle$ \\ \hline
$o_3$ & medium & $\langle 0.80,0.20 \rangle$ \\ \hline
\end{tabular}
\end{center}

\noindent
Given these observations and the rules, we can use the following algebraic
expression to identify the three objects:

\begin{center}
\begin{tabular}{l}
$\widehat{\pi}_{\mbox{Object-id,Object}}(\mbox{RadarData} ~\widehat{\bowtie}~ \mbox{RadarRules}) ~\widehat{\cap}~$ \\
$\widehat{\pi}_{\mbox{Object-id,Object}}(\mbox{GunData}  ~\widehat{\bowtie}~
\mbox{GunRules}) ~\widehat{\cap}~$ \\
$\widehat{\pi}_{\mbox{Object-id,Object}}(\mbox{SpeedData} ~\widehat{\bowtie}~
\mbox{SpeedRules})$
\end{tabular}
\end{center}

\noindent
The intuition behind the intersection is that we would like to 
capture the common (intersecting) information among the three
sensor data.
Evaluating this expression, we get the following paraconsistent relation:

\begin{center}
\begin{tabular}{|c|c||c|} \hline
Object-id & Object & Confidence Factors \\ \hline \hline
$o_1$ & T-72 & $\langle 0.05,0.0 \rangle$ \\ \hline
$o_2$ & T-80 & $\langle 0.0,0.05 \rangle$ \\ \hline
$o_3$ & T-80 & $\langle 0.05,0.0 \rangle$ \\ \hline
\end{tabular}
\end{center}

\noindent
It is clear from the result that by the given information,
we could not infer any useful information that is we could
not decide the status of objects $o_1, o_2$ and $o_3$.

\section{Conclusions and Future Work} \label{Conclusion}

We have presented a generalization of fuzzy relations, 
intuitionistic fuzzy relations (interval-valued fuzzy relations)
and paraconsistent relations,
called paraconsistent intuitionistic fuzzy relations,  in
which we allow the representation of confidence (belief and doubt)
factors with each tuple. The algebra on fuzzy relations is
appropriately generalized to manipulate paraconsistent intuitionistic fuzzy
relations. 

Various possibilities exist for further study in this area. 
Recently, there has been some work in extending logic programs
to involve quantitative paraconsistency. Paraconsistent logic programs
were introduced in \cite{blr89} and probabilistic logic programs in 
\cite{ngs92}. Paraconsistent logic programs 
allow negative atoms to appear in the
head of clauses (thereby resulting in the possibility of
dealing with inconsistency), and probabilistic logic programs associate 
confidence measures with literals and with entire clauses. 
The semantics of these
extensions of logic programs have already been presented, but
implementation strategies to answer queries have not been discussed.
We propose to use the model introduced in this paper 
in computing the
semantics of these extensions of logic programs.
Exploring application areas is another important thrust of
our research.  

We developed two notions of generalising operators on fuzzy  
relations
for paraconsistent intuitionistic fuzzy relations.
Of these, the stronger notion guarantees that any generalised  
operator is
``well-behaved'' for paraconsistent intuitionistic fuzzy relation 
operands that contain
consistent information.

For some well-known operators on fuzzy relations, such as union,  
join,
projection, we introduced generalised operators on paraconsistent 
intuitionistic fuzzy  
relations.
These generalised operators maintain the belief system intuition  
behind
paraconsistent intuitionistic fuzzy relations, and are shown to 
be ``well-behaved'' in the  
sense
mentioned above.

Our data model can be used to represent relational information that  
may be incomplete and inconsistent.
As usual, the algebraic operators can be used to construct queries to  
any database systems for retrieving vague information.

\section*{References}
\bibliography{ref}

\end{document}